\documentstyle[12pt,world_sci]{article}
\begin{document}

\title{{\bf FORMATION TIME SCALES FOR QUARKONIA IN A\\
DECONFINING MEDIUM}}
\author{R. L. THEWS\thanks{Supported by U. S. Department of Energy
Grant DE-FG02-85ER40213.}\\
{\em Department of Physics, University of Arizona, Tucson,\\
AZ 85721, USA}}

\maketitle
\setlength{\baselineskip}{2.6ex}

\begin{center}
\parbox{13.0cm}
{\begin{center} ABSTRACT \end{center}
{\small \hspace*{0.3cm}  Production of heavy quark antiquark systems in high
energy heavy ion collisions must involve relativistic momentum components in
a quantum mechanical approach.  If the color forces are screened in a
deconfining medium, one can define the analog of a formation or separation
time by an overlap integral in the nonrelativistic bound state rest frame.
This time parameter has some interesting properties which depend on the
momentum
spectrum of the initial quarks.  Consequences of these properties for the
phenomenology of deconfinement signals are discussed.}}
\end{center}

\section{Introduction}

The motivation for this study is  related to the possibility
of using suppression of heavy quarkonia states in high energy
heavy ion collisions as a signature for the formation of a quark-gluon
plasma.\cite{Matsui-Satz}
The heavy quark-antiquark pair is produced by hard collisions of
partons during the initial interaction times in the reactions.
If a quark-gluon plasma is formed, one would expect that the
confining color forces would be screened during the plasma
lifetime $t_p$ and hence the heavy quark-antiquark pair could
separate to a relative distance greater than that of the ordinary
bound states in the confining potential.
In the literature the minimum required time has been associated
wiht the
formation time of the bound state, but in this
context it is more properly understood as a separation time $t_s$
in a nonconfining screened potential.  When the confining potential
reappears, the quark-antiquark pair are separated too far to
fit into the confining potential region, so that at hadronization
they are most likely to recombine with ordinary light quarks, thus
leading to a suppression of the bound states with hidden flavor
content and an enhancement of the open flavor states.

\section{Classical Interpretation}

The observation of this predicted suppression in $J/\psi$ production
in O-U interactions by the NA38 collaboration at CERN \cite{NA38}
has lead to an avalanche of theoretical and phenomenological papers.
One feature of the data was immediately recognized as significant.
The suppression was maximal for small $J/\psi$ transverse momentum
$P_t$ and gradually disappeared at a critical value $P_c \approx M_{J/\psi}$
.  This is immediately understood in simple terms, since the
quark-antiquark system can avoid suppression if its separation time
(Lorentz dilated in the lab frame) is greater than either
the plasma lifetime $t_p$ or the time of transit of the pair
to the spatial boundary of the plasma region $t_b = xE/P_t$
where E is the transverse energy and x the transverse distance
to the boundary.  One must average over the production position
of the quark-antiquark pair, which leads directly to a linear
increase in the suppression factor. It reaches unity
at $P_c = {{Md}\over{t_s}}$, with d the transverse size of the plasma.
 If the plasma lifetime is the limiting parameter,
the linear rise will be truncated by an immediate saturation at a smaller
$P_c = M\sqrt{{{{t_p}^2}\over{{t_s}^2}}-1}$.
Of course, the
discontinuous values and slopes in this picture are artifacts
of the simple one-dimensional model.  Realistic calculations
\cite{Karsch} found good agreement with the data for
spatial parameters determined by the collision geometry, if
the plasma lifetime $t_p \approx 1 fm$ and the separation
time $t_s \approx 0.7 fm$.

\section{Quantum Mechanical Scenario}

These results seem to place quite severe constraints on the
parameters of a possible quark-gluon plasma. However, the
dynamics of formation of the quark-antiquark pair in the hard
collision tell a different story.  The dominant mechanism for
the production is gluon-gluon fusion, but any such process
is characterized by a scale set by the heavy quark or heavy
bound state mass.  Hence the spacetime region involved in the
production is  $\Delta x \sim {1\over {\Delta P}} \sim {1\over m_Q}$.
Since most of the events are produced  in a region $p \leq m_
{J/\psi} \approx 2m_Q$,
one has ${\Delta P \over {P}} \geq 1$
, i.e. we are in the quantum-mechanical regime where there are
not well-defined classical trajectories, and one cannot rely on
parameters extactly specifiying space and time events.
One notes in addition that since the mass scale is set by the bound
state or quark masses, one must also use relativistic momentum components
in the quantum-mechanical wave packet which describes the position
of the quarks.
The procedure is then straightforward:  one replaces the classical
trajectories of the quark and antiquark by expanding wave packets which
are initially localized in space and propagate with central momenta provided
in the production process.  The probability of this pair forming
the bound $J/\psi$ is just the overlap of the product of the wave
packets onto a superposition of bound state times total momentum
eigenstates.  If the wave packet propagation is in the normal
confining potential, these probabilities are time independent and
the ``formation time" plays no role.  However, if the color forces
are screened away as in a quark-gluon plasma, the evolution of the
wave packets is altered and the probability of production of the
$J/\psi$ becomes time-dependent.  This is where the quantum mechanical
analog of a separation time enters.

\section{A Simple Model}

We consider a one-dimensional example which can be done analytically.
The initial wave packets for quark and antiquark are Gaussian with
width $\sigma$ and initial (central) momenta $p_i$.  The bound state is
also Gaussian with width $\sigma_B$ and momentum P.  The packets are allowed
to propagate with mass m in a region of zero potential (simulating
the plasma phase) for a time t, and then the probability amplitude for
the formation of the bound state $a_{P}(t)$ is calculated:
\begin{equation}
  \label{eq:first}
  a_p (t) \sim e^{{\sigma^{2} \over 2} {(P-p_{1}-p_{2})^2 }}
  \int_{-\infty}^{\infty} {dk  e^{-it[(({1 \over 2} P + k)^2 + m^{2})^{1/2}
  + k\rightarrow -k]}} \ e^{-2\sigma^{2}(k- {1 \over 2}(p_{1} - p_{2}))^{2}}
  \ e^{-k^{2}\sigma_{B}^{2} }
\end{equation}

The suppression factor is the ratio of that probability  to
its value in the absence of the plasma phase.
One can define the quantum-mechanical separation time $t_s$ as the time
required for this ratio to decrease by a factor of 2, yielding
(for a non-relativistic bound state)
\begin{equation}
  \label{eq:second}
  t_{s}  =  \sqrt{3} m ( 2 \sigma^{2} + \sigma_{B}^{2} )
\end{equation}

Alternatively, one can use the time-dependent shape of this function
directly in the classical formulas for boundary crossing or
plasma lifetime constraints, suitably Lorentz dilated for $P \neq 0$.

This procedure has been followed both in this simple model and
for a full three-dimensional case with realistic collision
geometry \cite{QM}, and results in an effective classical separation
time ${t_s}^{QM} \approx 0.2 - 0.4 fm$.  According to the relationship
between the effective parameters and transverse-momentum cutoff values,
this then predicts $P_c$ values at least a factor of 2 or 3 greater
than allowed by experiment.  In retrospect, this is perhaps not so surprising,
since
the larger values of classical $t_s$ were estimated by average separation
momenta between quark and antiquark in the bound state potential, and
here we are allowing them to separate freely in zero relative potential
as would be the case for a completely screened confining force.

One must extend this analysis one more step, to study the situation
when the total momentum
of the bound state is relativistic.  Up to now we have
calculated the overlap integrals in the bound state rest frame, but if
it is moving relativistically in the lab frame where the wave packet
widths are defined, then one would expect that some Lorentz contraction
of the widths should be included.  A short examination of the formulas
involved reveals inconsistencies with this approach, and again it is
not surprising.  We have been dealing  with nonrelativistic quantum
mechanical amplitudes, and they of course do not transform as scalars,
but rather their squares transform as densities.  Hence one must
recast the momentum component expansion to exhibit this property:
\begin{equation}
  \label{eq:third}
  \varphi_{a} (x,t) = {1 \over {\sqrt{2\pi}}} \int {{dk} \over {\omega_k}}
  \sqrt{\omega_k} e^{i(kx- \omega_{k}t)} \Phi ( k \cdot a )
\end{equation}

In our example of Gaussian wave packets in the lab frame, this leads
to
\begin{equation}
  \label{eq:fourth}
  \Phi (k \cdot a) = \bigg( {2 \over \pi} \bigg)^{1 \over 4}
  \sqrt{k \cdot a} e^{-[(k \cdot a)^2 -
  m^2
  x^2_0] }
\end{equation}
where the 4-vector $a = (x_o,0)$ contains the width $x_o$ of the packet
in its rest frame.
In a moving frame, one transforms to $a^\prime = (\gamma x_o, \gamma \beta
 x_o)$
which for nonrelativistic momenta does in fact lead to the expected
Lorentz contraction of the wave packet.  However, the general case
leads to a much richer structure.

The probability amplitude in the
bound state rest frame becomes
\begin{equation}
  \label{eq:fifth}
  \begin{array}{l}
  a_p (t) \sim \int_{-\infty}^{\infty} dz \bigg( 1+ {s_1 \over c_1}
  {z \over {\sqrt {1+z^2}}} \bigg)^{1 \over 2}
  \bigg( 1- {s_2 \over c_2}
  {z \over {\sqrt {1+z^2}}} \bigg)^{1 \over 2}
  e^{-2 imt \sqrt{1+z^2}}
  \\ \times e^{-(m x_B)^2 z^2} \ e^{-(m x_0)^2 [s^2_1 + s^2_2 + z^2
(c^2_1+s^2_1+
  c^2_2 + s^2_2) +2(c_1 s_1 - c_2 s_2) z \sqrt{1+z^2}]}
\end{array}
\end{equation}
where we have used initial lab wave packet momenta $p_a$ and $p_b$, wave
packet widths $x_o$ and bound state width $x_B$ in their respective
rest frames, and
\begin{equation}
  \label{eq:sixth}
  \begin{array}{l}
  s = -{P \over {2m}}, s_a = {p_a \over m}, s_b = {p_b \over m}, \\
  s_1 = c s_a + s c_a, s_2 = c s_b + s c_b, c_i = \sqrt{1+ s^2_i}
\end{array}
\end{equation}
are the approprate transformation factors.  The integrals can be performed
in the case of a nonrelativistic bound state  $mx_B \gg 1$,
\begin{equation}
  \label{eq:seventh}
  \left| a_p (t) \right|^2 = {const \over { [\sigma^2 + (mt)^2]^{1 \over 2}}}
  \ \ e^{ \big( {{2 \sigma b^2} \over {\sigma^2 + (mt)^2} } \big) }
\end{equation}
where
\begin{equation}
  \label{eq:eighth}
  \begin{array}{l}
  \sigma \equiv (m x_B)^2 + 2 (mx_0)^2 ( 1+ 2 s_1^2 + 2 s^2_2) \\
  b \equiv (c_1 s_1 - c_2 s_2) (m x_0)^2
\end{array}
\end{equation}
One finds that the effective separation time ${t_s}^{QM} = t_s(P)$, i.e. the
bound state ``remembers"
the momenta of the quark pair which led to its formation.  For the
situation $p_a = p_b$ (where the hard production amplitude is
maximum), one finds
\begin{equation}
  \label{ninth}
  t_s^{QM} = {\sqrt 3 \over m} \big[ (m x_B)^2 + (m x_0)^2
  (2 + {P^2 \over m^2}) \big]
\end{equation}

Again, this is the effective separation time {\it in the rest frame of
the bound state},  and the P-dependence is in addition to that
which will occur in the Lorentz dilation transformation to the lab frame.

Note that the direction of this factor is to increase the separation time
as a function of bound state momentum, i.e. to bring the quantum-mechanical
parameters back into  a region which could be compatible with experiment.
For example, the parameters in Eq. 9 would lead to a reduction in
$P_c$ from 5 GeV/c down to less than 3 GeV/c.  No fit to the data
is attempted here, since we are dealing only with a one-dimensional
model.

As mentioned previously, one must average over production position
in a realistic nuclear geometry to get the final results.  If we
simulate this situation here with a uniform density over a transverse
size which yields the above $P_c$'s, one finds a considerable flattening
of the suppression curves.  This effect tends to oppose the desirable
results of the separation time increase, but in principle the total
result will still be reflected in the data.

An opposite point of view may also be examined in this context.  There
is a possibility that initial state effects in nuclear matter can
mimic the $P_t$ dependence of suppression by skewing the transverse
momentum distributions of the incoming partons which participate in
the hard collisions.  This scenario has been examined \cite{initial}
and under certain assumptions could possibly account for the entire
effect observed in the NA38 data.  If this is the actual situation,
one can use the results we have developed to put constraints on the
plasma parameters such that it will not induce  $P_t$
dependence in excess of that in the data.  In our simple model, this
can easily be seen as a requirement on the plasma lifetime $t_p \geq
\sqrt{2}  t_s$, and also on the transverse size of the plasma  $d \gg t_s$.

\section{Summary}

The primary result of this investigation is that production of
heavy quarkonium states in hadronic interactions must involve
momentum components for the initially-produced quarks which are both
relativistic and have an uncertainty which precludes using
classical trajectories for their description.  When applied to
scenarios in which color forces are temporarily screened, such as
 in a quark gluon-plasma, one is led to a quantum mechanical
version of a ``formation time" (in reality it is a separation time),
which must be calculated from overlap integrals of quantum mechanical
wave packets and bound states.  This has a significant
effect on the $P_t$ - dependence of suppression of quarkonium
formation as interpreted as a signal for quark-gluon plasma
formation.  One unanticipated feature is the effect of the initial
hard production process momentum dependence on the separation time,
even when calculated in the bound state rest frame.  It is shown that
this additional effect tends to sharpen the cut-off slope of the
suppression vs. $P_t$ curve. Averaging over nuclear
geometry  in the production process has a
tendency to reduce its magnitude.  Some remnants
of this effect will
be present in any scenario of plasma formation, even
if initial state nuclear effects are responsible for the
presently-observed $P_t$ dependence in the data. Thus  one
can derive constraints on plasma parameters from either the observation
or non-observation of this effect.

\bibliographystyle{unsrt}

\end{document}